\documentclass[a4paper,11pt]{article}
\usepackage{jinstpub} 
\usepackage{booktabs}
\usepackage[british]{babel}
\usepackage{xspace}
\usepackage{subcaption}
\usepackage{bm}

\usepackage[version=4]{mhchem} 
\ExplSyntaxOn
\keys_define:nn { mhchem }
 {
  arrow-min-length .code:n =
   \cs_set:Npn \__mhchem_arrow_options_minLength:n { {#1} } 
 }
\ExplSyntaxOff
\mhchemoptions{arrow-min-length=1.5em}

\usepackage[separate-uncertainty, alsoload=synchem, detect-weight=true]{siunitx}

\newcommand{\cevns}{CEvNS\xspace}

\title{A \ce{D2O} detector for flux normalization of a pion decay-at-rest neutrino source}

\newcommand{\Mephi}{a}
\newcommand{\Mephidesc}{\affiliation[\Mephi]{National Research Nuclear University MEPhI (Moscow Engineering Physics Institute), Moscow, 115409, Russian Federation}}
\newcommand{\Duke}{b}
\newcommand{\Dukedesc}{\affiliation[\Duke]{Department of Physics, Duke University, Durham, NC, 27708, USA}}
\newcommand{\TUNL}{c}
\newcommand{\TUNLdesc}{\affiliation[\TUNL]{Triangle Universities Nuclear Laboratory, Durham, NC, 27708, USA}}
\newcommand{\UTK}{d}
\newcommand{\UTKdesc}{\affiliation[\UTK]{Department of Physics and Astronomy, University of Tennessee, Knoxville, TN, 37996, USA}}
\newcommand{\ITEP}{e}
\newcommand{\ITEPdesc}{\affiliation[\ITEP]{Institute for Theoretical and Experimental Physics named by A.I. Alikhanov of National Research Centre ``Kurchatov Institute'', Moscow, 117218, Russian Federation}}
\newcommand{\ORNL}{f}
\newcommand{\ORNLdesc}{\affiliation[\ORNL]{Oak Ridge National Laboratory, Oak Ridge, TN, 37831, USA}}
\newcommand{\Sandia}{g}
\newcommand{\Sandiadesc}{\affiliation[\Sandia]{Sandia National Laboratories, Livermore, CA, 94550, USA}}
\newcommand{\USD}{h}
\newcommand{\USDdesc}{\affiliation[\USD]{Physics Department, University of South Dakota, Vermillion, SD, 57069, USA}}
\newcommand{\CMU}{i}
\newcommand{\CMUdesc}{\affiliation[\CMU]{Department of Physics, Carnegie Mellon University, Pittsburgh, PA, 15213, USA}}
\newcommand{\UW}{j}
\newcommand{\UWdesc}{\affiliation[\UW]{Center for Experimental Nuclear Physics and Astrophysics \& Department of Physics, University of Washington, Seattle, WA, 98195, USA}}
\newcommand{\LANL}{k}
\newcommand{\LANLdesc}{\affiliation[\LANL]{Los Alamos National Laboratory, Los Alamos, NM, 87545, USA}}
\newcommand{\Laurentian}{l}
\newcommand{\Laurentiandesc}{\affiliation[\Laurentian]{Department of Physics, Laurentian University, Sudbury, Ontario, P3E 2C6, Canada}}
\newcommand{\NCSU}{m}
\newcommand{\NCSUdesc}{\affiliation[\NCSU]{Department of Physics, North Carolina State University, Raleigh, NC, 27695, USA}}
\newcommand{\IU}{n}
\newcommand{\IUdesc}{\affiliation[\IU]{Department of Physics, Indiana University, Bloomington, IN, 47405, USA}}
\newcommand{\VT}{o}
\newcommand{\VTdesc}{\affiliation[\VT]{Center for Neutrino Physics, Virginia Tech, Blacksburg, VA, 24061, USA}}
\newcommand{\NCCU}{p}
\newcommand{\NCCUdesc}{\affiliation[\NCCU]{Department of Mathematics and Physics, North Carolina Central University, Durham, NC, 27707, USA}}
\newcommand{\UF}{q}
\newcommand{\UFdesc}{\affiliation[\UF]{Department of Physics, University of Florida, Gainesville, FL, 32611, USA}}
\newcommand{\Tufts}{r}
\newcommand{\Tuftsdesc}{\affiliation[\Tufts]{Department of Physics and Astronomy, Tufts University, Medford, MA, 02155, USA}}
\newcommand{\KAIST}{s}
\newcommand{\KAISTdesc}{\affiliation[\KAIST]{Department of Physics, Korea Advanced Institute of Science and Technology, Daejeon, 34141, Republic of Korea}}
\newcommand{\CAPP}{t}
\newcommand{\CAPPdesc}{\affiliation[\CAPP]{Center for Axion and Precision Physics Research (CAPP) at Institute for Basic Science (IBS), Daejeon, 34141, Republic of Korea}}
\author[\Mephi]{D. Akimov,}\Mephidesc
\author[\Duke,\TUNL]{P. An,}\Dukedesc\TUNLdesc
\author[\Duke,\TUNL]{C. Awe,}
\author[\Duke,\TUNL]{P.S. Barbeau,}
\author[\UTK]{B. Becker,}\UTKdesc
\author[\ITEP,\Mephi]{V. Belov ,}\ITEPdesc
\author[\UTK]{I. Bernardi,}
\author[\ORNL]{M.A. Blackston,}\ORNLdesc
\author[\Mephi]{A. Bolozdynya,}
\author[\Sandia]{B. Cabrera-Palmer,}\Sandiadesc
\author[\USD]{D. Chernyak,}\USDdesc
\author[\Duke]{E. Conley,}
\author[\UTK,\ORNL]{J. Daughhetee,}
\author[\CMU]{E. Day,}\CMUdesc
\author[\UW]{J. Detwiler,}\UWdesc
\author[\USD]{K. Ding,}
\author[\UW]{M.R. Durand,}
\author[\UTK,\ORNL]{Y. Efremenko,}
\author[\LANL]{S.R. Elliott,}\LANLdesc
\author[\ORNL]{L. Fabris,}
\author[\ORNL]{M. Febbraro,}
\author[\Laurentian]{A. Gallo Rosso,}\Laurentiandesc
\author[\ORNL,\UTK]{A. Galindo-Uribarri,}
\author[\TUNL,\ORNL,\NCSU]{M.P. Green ,}\NCSUdesc
\author[\ORNL]{M.R. Heath,}
\author[\Duke,\TUNL]{S. Hedges,}
\author[\CMU]{D. Hoang,}
\author[\IU]{M. Hughes,}\IUdesc
\author[\Duke,\TUNL]{T. Johnson,}
\author[\Mephi]{A. Khromov,}
\author[\Mephi,\ITEP]{A. Konovalov,}
\author[\TUNL,1]{J. Koros,}\note{Now at: Department of Physics, University of Notre Dame, IN, 46556, USA}
\author[\Mephi,\ITEP]{E. Kozlova,}
\author[\Mephi]{A. Kumpan,}
\author[\Duke,\TUNL]{L. Li,}
\author[\VT]{J.M. Link,}\VTdesc
\author[\USD]{J. Liu,}
\author[\NCSU]{K. Mann,}
\author[\NCCU,\TUNL]{D.M. Markoff,}\NCCUdesc
\author[\IU]{J. Mastroberti,}
\author[\ORNL]{P.E. Mueller,}
\author[\ORNL]{J. Newby,}
\author[\CMU]{D.S. Parno,}
\author[\ORNL]{S.I. Penttila,}
\author[\Duke]{D. Pershey,}
\author[\CMU]{R. Rapp,}
\author[\UF]{H. Ray,}\UFdesc
\author[\Duke]{J. Raybern,}
\author[\Mephi,\ITEP]{O. Razuvaeva,}
\author[\Sandia]{D. Reyna,}
\author[\TUNL]{G.C. Rich,}
\author[\NCCU,\TUNL]{J. Ross,}
\author[\Mephi]{D. Rudik,}
\author[\Duke,\TUNL]{J. Runge,}
\author[\IU]{D.J. Salvat,}
\author[\CMU]{A.M. Salyapongse,}
\author[\Duke]{K. Scholberg,}
\author[\Mephi]{A. Shakirov,}
\author[\Mephi,\ITEP]{G. Simakov,}
\author[\Duke,2]{G. Sinev,}\note{Now at: South Dakota School of Mines and Technology, Rapid City, SD, 57701, USA}
\author[\IU]{W.M. Snow,}
\author[\Mephi]{V. Sosnovstsev,}
\author[\IU]{B. Suh,}
\author[\IU]{R. Tayloe,}
\author[\VT]{K. Tellez-Giron-Flores,}
\author[\IU,3]{I. Tolstukhin,}\note{Now at: Argonne National Laboratory, Argonne, IL, 60439, USA}
\author[\NCCU,\TUNL]{E. Ujah,}
\author[\IU]{J. Vanderwerp,}
\author[\ORNL]{R.L. Varner,}
\author[\Laurentian]{C.J. Virtue,}
\author[\IU]{G. Visser,}
\author[\UTK]{E.M. Ward,}
\author[\UW]{C. Wiseman,}
\author[\Tufts]{T. Wongjirad,}\Tuftsdesc
\author[\CMU]{Y.-R. Yen,}
\author[\KAIST,\CAPP]{J. Yoo,}\KAISTdesc\CAPPdesc
\author[\ORNL]{C.-H. Yu,}
\author[\IU,4]{J. Zettlemoyer}\note{Now at: Fermi National Accelerator Laboratory, Batavia, IL, 60510, USA}

\collaboration{COHERENT collaboration}

\emailAdd{heathmr@ornl.gov}

\newpage
\abstract{
    We report on the technical design and expected performance of a \SI{592}{\kg} heavy-water-Cherenkov detector to measure the absolute neutrino flux from the pion-decay-at-rest neutrino source at the Spallation Neutron Source (SNS) at Oak Ridge National Laboratory (ORNL). 
    The detector will be located roughly \SI{20}{\m} from the SNS target and will measure the neutrino flux with better than \SI{5}{\percent} statistical uncertainty in \SI{2}{years}.
    This heavy-water detector will serve as the first module of a two-module detector system to ultimately measure the neutrino flux to \SIrange[range-phrase=--, range-units=single]{2}{3}{\percent} at both the First Target Station and the planned Second Target Station of the SNS.
    This detector will significantly reduce a dominant systematic uncertainty for neutrino cross-section measurements at the SNS, increasing the sensitivity of searches for new physics.
}


\arxivnumber{2104.09605} 

\begin{document}

\maketitle
\footnotetext[5]{
This manuscript has been authored by UT-Battelle, LLC, under contract DE-AC05-00OR22725 with the US Department of Energy (DOE).
The US government retains and the publisher, by accepting the article for publication, acknowledges that the US government retains a nonexclusive, paid-up, irrevocable, worldwide license to publish or reproduce the published form of this manuscript, or allow others to do so, for US government purposes.
DOE will provide public access to these results of federally sponsored research in accordance with the DOE Public Access Plan \url{(http://energy.gov/downloads/doe-public-access-plan)}.
}

\section{Introduction}
\label{sec:intro}

Coherent elastic neutrino-nucleus scattering (\cevns) is a type of neutral-current neutrino scattering that was predicted in 1974~\cite{Freedman:1973yd,Kopeliovich:1974mv} and experimentally observed in 2017~\cite{Akimov:2017ade} by the COHERENT experiment. 
In \cevns, a neutrino scatters coherently from an entire nucleus rather than from an individual nucleon, resulting in a large boost to the cross section. 
With a cross section on the order of \SI{1e-39}{\cm\squared}, the \cevns interaction is roughly 100~times more likely to occur than other neutrino interactions at a few tens of \si{\MeV} for heavy nuclei, so \cevns-sensitive neutrino detectors can be made quite compact. 
Unfortunately, the \cevns process is extremely challenging to detect because of the tiny nuclear recoil energies involved. 
Nonetheless, COHERENT's initial result on \ce{CsI} was followed by a positive observation on argon~\cite{Akimov:2020pdx}.

The Standard Model provides a very clean and direct prediction of the \cevns cross section~\cite{Barranco_2005}. 
A precise measurement of the \cevns cross section can thus be used to search for new physics beyond the Standard Model. 
For example, the neutrino magnetic moment~\cite{VogelEngel:1989, Dodd:1991ni, Scholberg:2005qs, Kosmas:2015} and the neutrino charge radius~\cite{Papavassiliou:2005cs} will affect the \cevns cross section.
COHERENT's first \cevns measurement has already been used to set competitive bounds on the neutrino charge radius~\cite{Cadeddu-ChargeRadius:2018}. 
Since it is a neutral-current process, \cevns also provides an attractive possible strategy for an oscillation-based sterile-neutrino search~\cite{formaggio:2012, anderson:2012}.
Finally, the \cevns cross section is also sensitive to so-called non-standard neutrino interactions (NSI)~\cite{Barranco:2005yy, Scholberg:2005qs, Dutta:2015vwa, Miranda:2020tif, Coloma:2019mbs}, beyond-Standard Model interactions of neutrinos and quarks.
Through this sensitivity to NSI, \cevns measurements can help to resolve degeneracies in the mass-ordering determination from long-baseline neutrino experiments such as DUNE~\cite{ColomaSchwetz-MassOrderingDegeneracy:2016, coloma:2017, Coloma-LMADark:2017, 2020JHEP...09..106D}.

Even in the absence of new physics, \cevns measurements will help untangle the low-energy nuclear recoil responses of various detector technologies, providing valuable calibration input for WIMP dark-matter searches with similar technologies. 
As the interaction and detector response are better characterized, \cevns will become a possible tool for compact nuclear nonproliferation monitors. Unlike inverse beta decay, it is sensitive to low-energy capture antineutrinos from uranium breeder blankets~\cite{CogswellHuber-NuclearMonitoring:2016}, as \cevns is sensitive to neutrinos below the inverse beta decay threshold.
Finally, precision measurements of \cevns cross sections will improve our understanding of the target nuclei through measurements of the neutron radius~\cite{Cadeddu:2021ijh, Caddedu-CsINeutronDensity-2018, Amanik_2008, Patton-NeutronDensity-2012}, which is a crucial input for our understanding of neutron-rich matter, from exotic nuclei to neutron stars.

The COHERENT experiment is located in ``Neutrino Alley,'' a basement corridor roughly \SI{20}{\m} from the pion-decay-at-rest neutrino source at the Spallation Neutron Source (SNS), a US Department of Energy (DOE) User Facility located at Oak Ridge National Laboratory (ORNL).
COHERENT is establishing a long-term program to study the CEvNS interaction on various nuclei to confirm the proportionality of the cross section to the number of neutrons squared ($N^{2}$), and to use CEvNS for some of the physics measurements mentioned above. 
In addition to the \cevns measurements with \ce{CsI} and argon made to date~\cite{Akimov:2017ade, Akimov:2020pdx}, the collaboration is deploying additional detectors to search for \cevns in multiple materials (\ce{NaI} and germanium) to further test the $N^{2}$ dependence, as well as measuring the cross sections of other neutrino interactions of particular relevance to neutrino and supernova physics~\cite{Akimov:2018ghi}. 
For example, the charged-current cross section \ce{{}^{127}I(\nu_{e},e^{-}){}^{127}Xe^{*}} is under study, along with neutrino-induced neutron emission of lead and iron.
In addition, charged-current neutrino interactions with argon, relevant for interpreting a DUNE supernova signature~\cite{2020arXiv200806647D}, will be studied with a tonne-scale liquid argon detector in the future~\cite{Akimov:2018ghi}.

For the current COHERENT results~\cite{Akimov:2017ade, Akimov:2020pdx}, one of the dominant systematic uncertainties is due to the estimated \SI{10}{\percent} uncertainty on the neutrino flux from the SNS target.
This uncertainty arises from comparisons between model predictions and the sparse available world data.
Surface-based Cherenkov detectors have been successfully operated in the past, see e.g.~\cite{PhysRevLett.44.522, Aguilar-Arevalo:2020nvw, Back:2017kfo}.
To that end, a \SI{592}{\kg} heavy-water Cherenkov detector has been designed to operate in Neutrino Alley.
This detector will make use of the well-understood \ce{\nu_{e} + d} interaction cross section~\cite{PhysRevC.75.044610, PhysRevC.101.054001, PhysRevC.101.015505} to greatly reduce the uncertainty on the SNS neutrino flux.

This paper describes the design and expected performance of such a heavy-water detector.
Section~\ref{sec:nusource} describes the SNS neutrino source.
Section~\ref{sec:nu-d2o} describes the current status of the relevant theoretical \ce{D2O} neutrino cross sections.
Section~\ref{sec:design-fundamentals} describes the design of the detector.
Section~\ref{sec:performance} covers the expected detector performance.
Sections~\ref{sec:sig_and_bkg} and \ref{sec:rates} cover the expected signals and backgrounds and their predicted rates.
Finally, Section~\ref{sec:phys_impact} discusses the expected physics impacts of such a heavy-water detector operating at the SNS.

\section{Neutrinos from the Spallation Neutron Source}
\label{sec:nusource}

\begin{figure}
    \centerline{
        \includegraphics[width=0.495\textwidth]{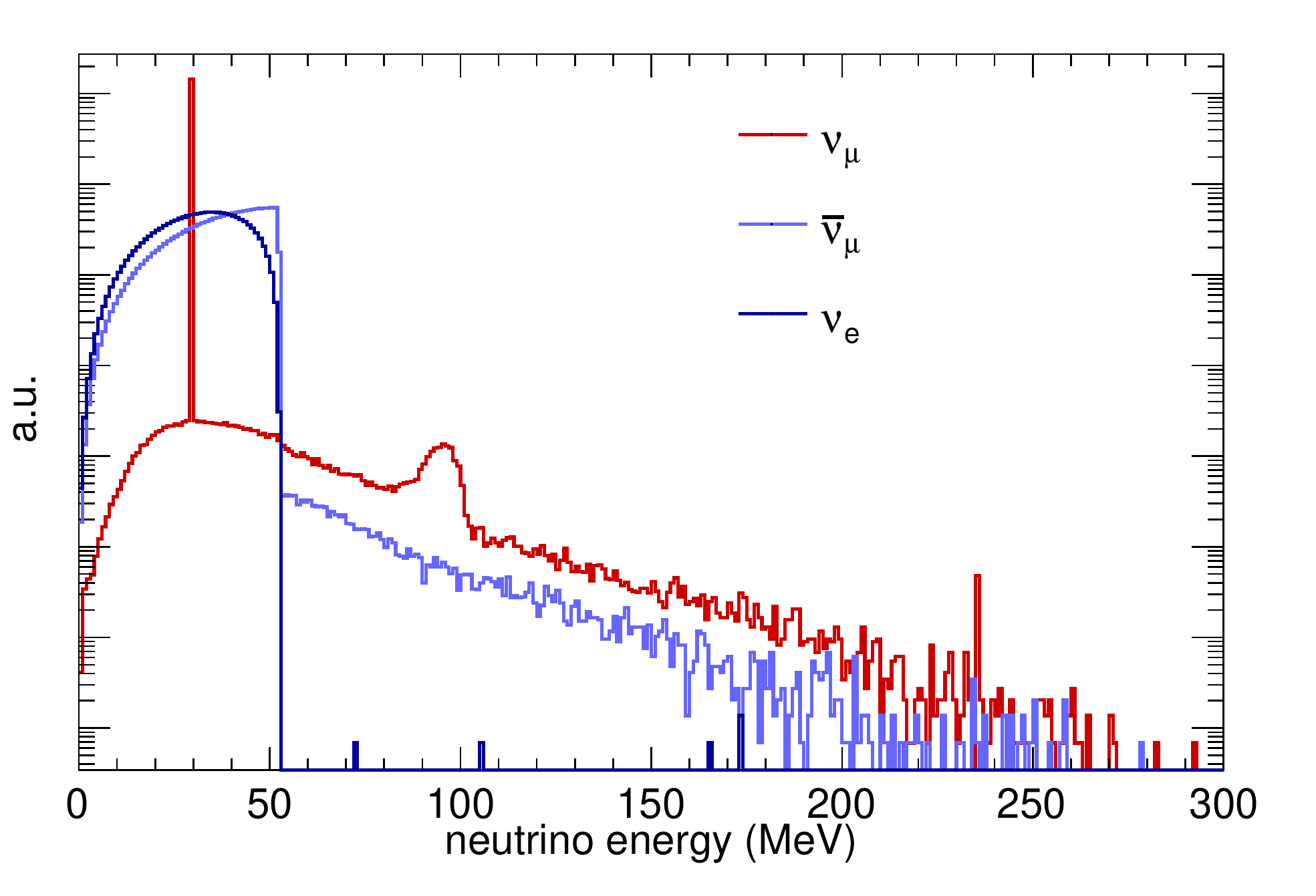}
        \includegraphics[width=0.495\textwidth]{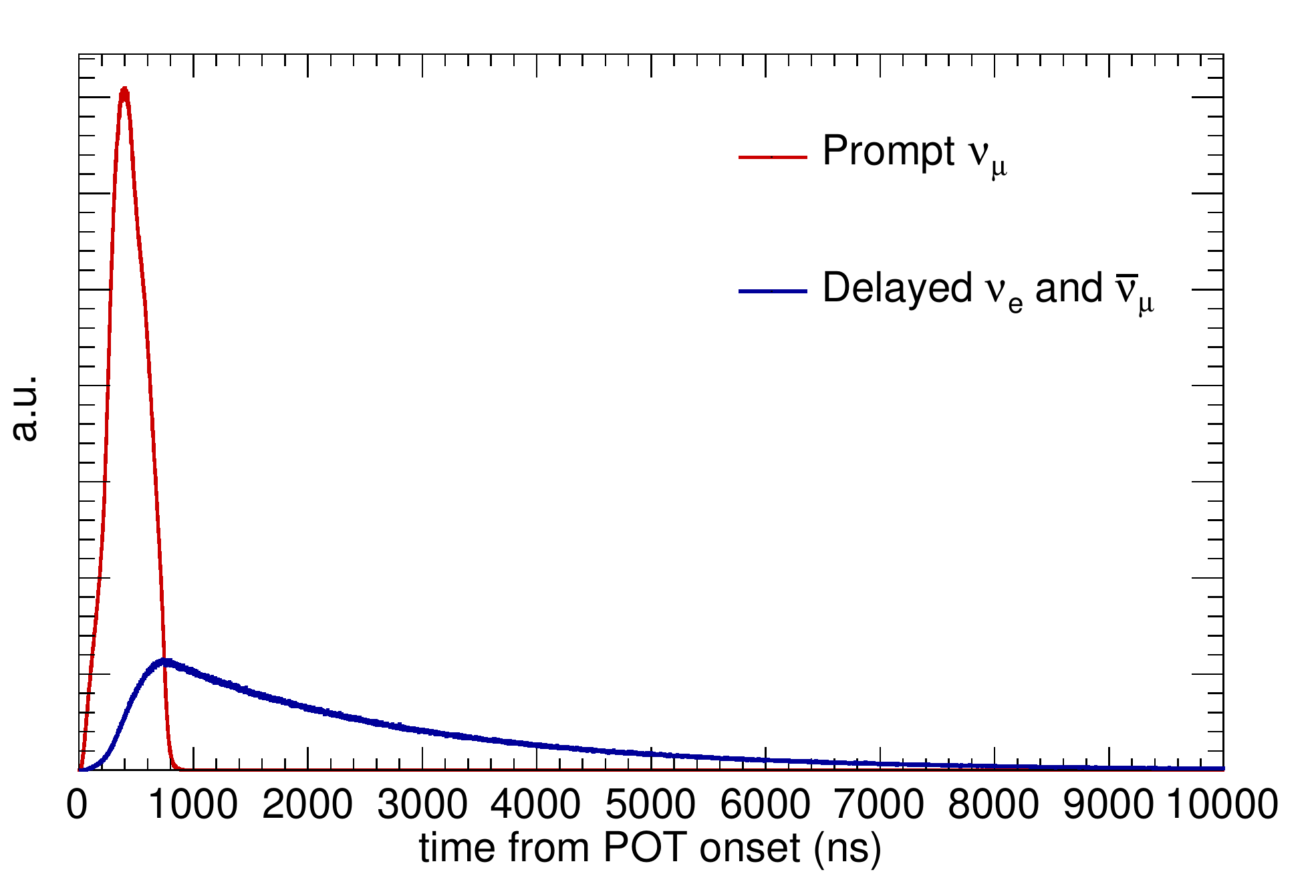}
    }
    \caption{Simulated energy (left) and time (right) spectra for neutrinos from a stopped-pion neutrino source like the ORNL SNS.}
    \label{fig:e_t_spect}
\end{figure}

The SNS is a user facility that operates a \SI{1.4}{\MW} proton beam. 
Neutrons at the SNS are produced by bunches of roughly \SI{1}{\GeV} protons impinging on a thick liquid-mercury target.
In addition to the neutrons produced at the SNS, copious charged pions are also produced in the target, with approximately \num{0.09} $\pi^{+}$ created for every proton on target (POT)~\cite{Rapp:2019vnv}.
Approximately \SI{99}{\percent} of the $\pi^{-}$ are captured in the thick mercury target, while the $\pi^{+}$ stop and decay at rest in the target.
These decaying pions ultimately produce three different neutrinos for each pion decay, with an energy spectrum extending up to \SI{53}{\MeV} (Figure~\ref{fig:e_t_spect}):
\begin{equation}
\pi^+ \rightarrow \nu_\mu + \mu^+ ; \qquad
\mu^+ \rightarrow \bar{\nu}_\mu + \nu_e + e^{+}. 
\end{equation}
This results in an approximate neutrino flux of \SI{4.3e7}{neutrinos\per\cm\squared\per\s} at a distance of \SI{20}{\m}~\cite{Akimov:2018ghi}.
Owing to the short $\pi^{+}$ lifetime (\SI{26}{\ns}), the $\nu_{\mu}$ are produced roughly in tandem with the beam pulse and follow the protons-on-target time profile.
The $\bar{\nu}_{\mu}$ and $\nu_{e}$ are delayed according to the lifetime of the $\mu^{+}$ decay (\SI{2.2}{\micro\s}) as seen in Figure~\ref{fig:e_t_spect}.

These neutrinos are ideal for a \cevns experiment.
The relatively large neutrino energies produce larger nuclear-recoil energies than do reactor-produced neutrinos, although very low detector thresholds are still required. 
In addition, because of the underlying timing structure of the SNS proton beam (\SI{60}{\Hz} bursts with full width at half maximum of \SI{350}{\ns} each), neutrinos arrive at the COHERENT detectors in well-defined pulses.
This timing permits neutrino-flavor separation as well as the precise subtraction of steady-state backgrounds. 

The neutrino flux from the SNS target was estimated with a  \textsc{Geant4}~\cite{AGOSTINELLI2003250, ALLISON2016186, Allison:2006ve} simulation, which relies on the Bertini intranuclear cascade model of hadronic interactions~\cite{PhysRev.188.1711} to predict the intensity of pion production in the target based on the characteristics of the proton beam~\cite{Rapp:2019vnv}. 
However, the pion-production cross section for \SI{1}{\GeV} protons on a mercury target has never been measured. 
Furthermore, a full accounting of charged-pion production in the target requires knowledge of pion-production cross-sections at lower energies, as protons rapidly lose energy to ionization interactions once they enter the SNS target, and at all production angles. 
Uncertainties in the model predictions give rise to an estimated \SI{10}{\percent} uncertainty on the neutrino flux, the second-largest systematic in COHERENT's published result on \ce{CsI}~\cite{Akimov:2017ade} and the largest on the full \ce{CsI} result, following an improved determination of the \ce{CsI} quenching factor by the COHERENT collaboration~\cite{magcevns_dan_2020, magcevns_alexey_2020}. 
It is also the largest systematic uncertainty in our first result on argon~\cite{Akimov:2020pdx} and is the dominant systematic uncertainty shared across all COHERENT neutrino detectors.
Reducing the neutrino-flux uncertainty will benefit cross-section measurements, and the corresponding physics impact, for all COHERENT detectors.

To reduce this systematic, a \SI{592}{\kg} heavy-water detector has been designed, taking advantage of the current \SIrange[range-phrase=--, range-units=single]{2}{3}{\percent} theoretical uncertainty (Section~\ref{sec:nu-d2o}) on the charged-current deuteron cross section to measure the neutrino flux directly.

\section{Neutrino Interactions on \ce{D2O}}
\label{sec:nu-d2o}
Many neutrino-interaction cross sections of nuclei in the energy range relevant for pion-decay-at-rest neutrinos (a few tens of \si{\MeV}) are not well constrained experimentally~\cite{RevModPhys.84.1307}. 
The charged-current cross section on deuterium is one of the best-understood interactions~\cite{PhysRevLett.43.96, PhysRevLett.44.522, 1991JETPL..53..513V, PhysRevC.59.1780, kozlov:2000, kozlov:2002}. 
This cross section has also benefited from extensive theoretical studies using a variety of tools, including potential models, pionless effective field theory, and the EFT* approach of combining potential-model wave functions with transition operators from chiral perturbation theory~\cite{RevModPhys.83.195}. 

Based on these theoretical studies, the \ce{\nu_{e} + d} interaction cross section is constrained to \SIrange[range-phrase=--, range-units=single]{2}{3}{\percent} simply by the agreement between unconstrained calculations using these approaches~\cite{PhysRevC.75.044610, PhysRevC.101.054001}.
This uncertainty envelope may be further reduced by experimental constraints on the axial exchange-current operator~\cite{RevModPhys.83.195,PhysRevC.101.054001,PhysRevC.44.619}, and active theoretical progress is continuing~\cite{PhysRevC.101.054001,PhysRevC.101.015505}. 
Sensitivity calculations are ongoing, but we anticipate that with \SI{592}{\kg} of \ce{D2O}, we can achieve a statistical precision of better than \SI{5}{\percent} in 2 years of running time at the SNS, as discussed below, approaching the current theoretical uncertainty of the \ce{\nu_{e} + d} cross section. 
\begin{figure}
\begin{center}
\includegraphics[width=0.65\textwidth]{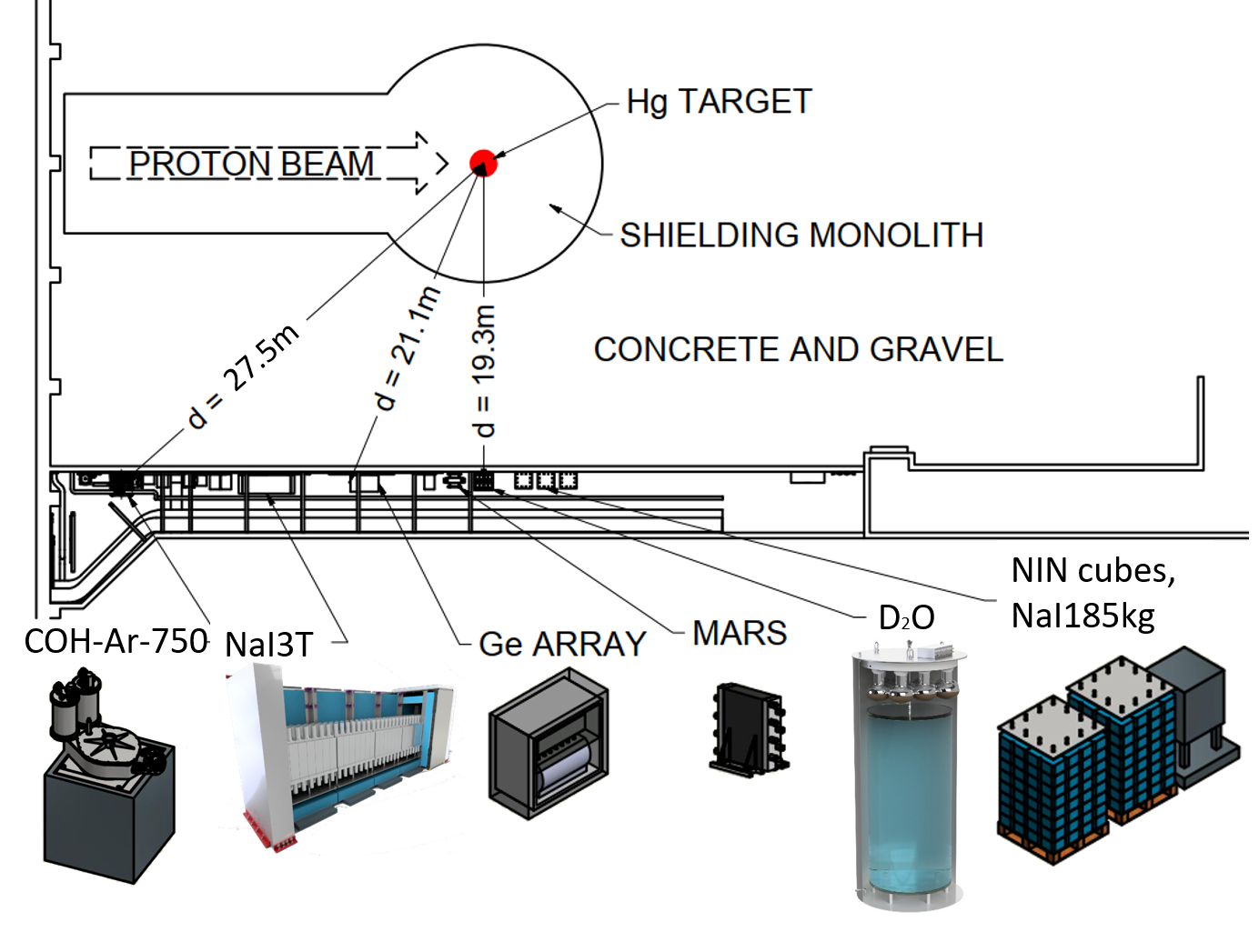}
\caption{Planned near-future deployment of COHERENT detectors in Neutrino Alley at the SNS. The precise siting of the Multiplicity And Recoil Spectrometer (MARS) neutron monitoring detector can readily be changed based on the needs of the collaboration.}
\label{fig:NuAlley}
\end{center}
\end{figure}

In addition to charged-current deuteron scattering, a heavy-water detector can also study the \ce{\nu_{e} + {}^{16}O -> e^{-} + {}^{16}F^{*}} charged-current interaction. 
Such a measurement would provide valuable cross-section information for supernova detection in existing and future large water Cherenkov detectors~\cite{PhysRevD.97.072001, Abe:2018uyc}. 

\section{Fundamentals of \ce{D2O} Detector Design} 
\label{sec:design-fundamentals}

A \SI{592}{\kg} heavy-water detector has been designed to fit alongside the various neutrino detectors present at the SNS in Neutrino Alley (Figure~\ref{fig:NuAlley}).
It will both provide an initial measurement of the neutrino flux from the SNS and study beam-related backgrounds, light-collection efficiency, and event reconstruction in situ. 
The detector will be externally triggered on SNS timing signals during SNS operations giving a threshold-less trigger.
An analogous trigger will be used to trigger the data acquisition in between SNS beam pulses allowing for an in situ unbiased measurement of beam-unrelated backgrounds.
Note that the heavy-water detector will operate in a different energy regime from \cevns detectors in Neutrino Alley, searching for tens-of-\si{\MeV} electrons.
This detector will serve as the first stage of a two-module detector that will reduce one of the leading COHERENT uncertainties by a factor of 2 or better. 
Improving this flux-normalization uncertainty is a critical milestone on the path to precision \cevns measurements at the SNS. 

\begin{figure}
     \centering
     \begin{subfigure}[b]{0.20\textwidth}
         \centering
         \includegraphics[width=\textwidth]{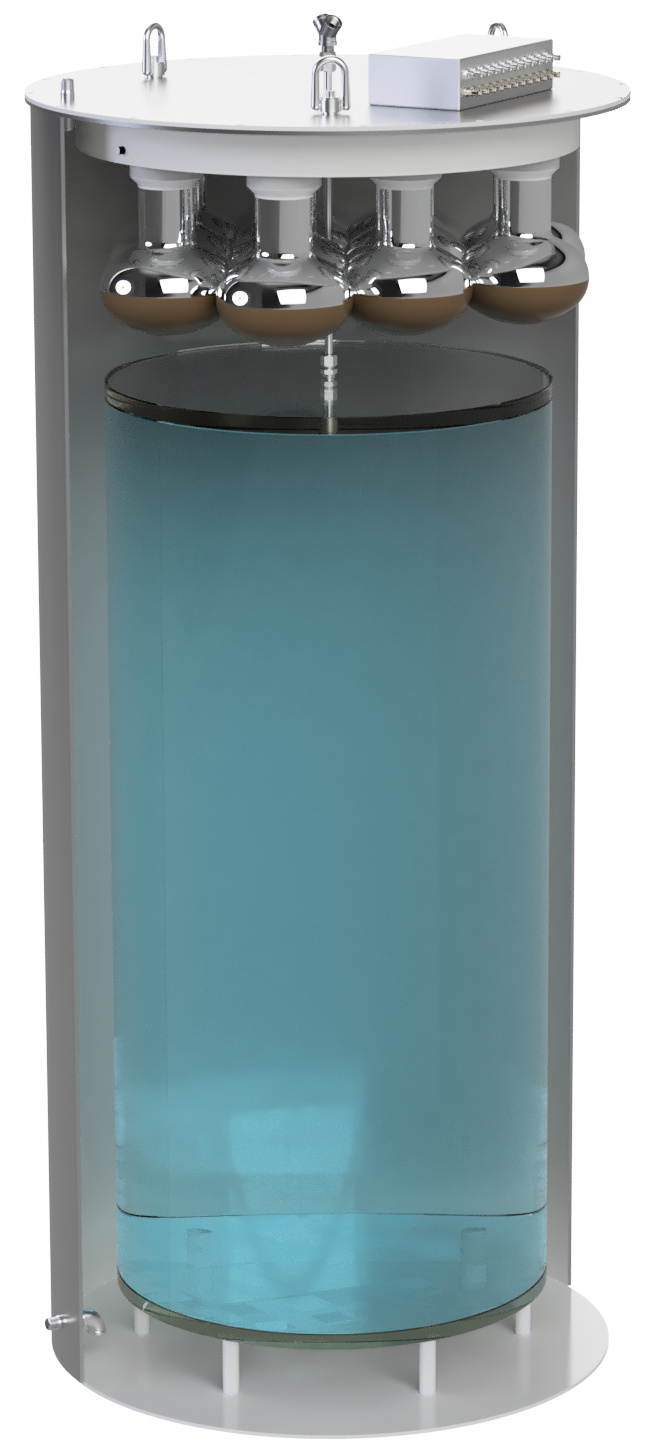}
		\label{fig:d20-cad}    
 	\end{subfigure}
 	\hspace*{0.06\textwidth}
    \begin{subfigure}[b]{0.22\textwidth}
         \centering
         \includegraphics[width=\textwidth]{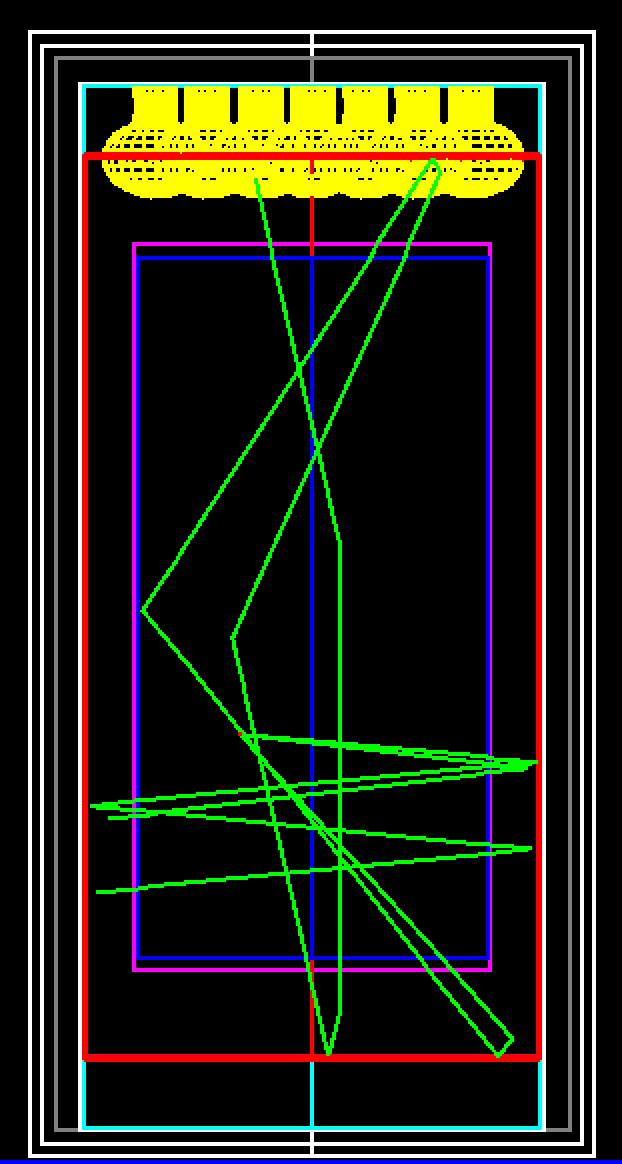}
		\label{fig:d20-design}    
 	\end{subfigure}
        \caption{\textit{Left:} Engineering drawing of the detector, showing the fiducial volume of \ce{D2O} in blue, along with the photomultiplier tubes, inside the steel tank. \textit{Right:} Side view of a simulated event. A small number of Cherenkov-photon paths generated by an electron are shown in green. Also visible is the Tyvek\textsuperscript{\textregistered} reflector (red), steel vessel (cyan), \ce{Pb} shielding (gray), and two-layer muon veto (white).}
        \label{fig:d2o-basics}
\end{figure}

The \ce{D2O} detector will be deployed roughly \ang{90} off-axis and approximately \SI{20}{\m} from the SNS target (Figure~\ref{fig:NuAlley}). 
At this location, there is very little contribution from decay-in-flight neutrinos (less than \SI{1}{\percent} of the total neutrino spectrum), which are boosted forward. 
Since Neutrino Alley is a basement corridor with required egress routes, there is no space for photomultiplier tubes (PMTs) to be mounted perpendicular to the hallway walls.
Even in that orientation, the space constraints in Neutrino Alley, along with passive and active shielding requirements for the detector, limit the possible width of the detector.
To satisfy these constraints with a cost-effective system, the detector geometry consists of an upright cylinder, with \SI{592}{\kg} of heavy water contained inside a clear acrylic vessel \SI{70}{\cm} in diameter and \SI{140}{\cm} tall (Figure~\ref{fig:d2o-basics}). 
The acrylic vessel is \SI{0.6}{\cm} thick on the sides with \SI{2.54}{\cm} thick end caps.

Within this well-defined volume, electrons from the primary reaction
\begin{equation}
    \ce{\nu_e + d -> p + p + e^{-}}
\end{equation}
produce Cherenkov radiation, which passes through the acrylic container into an outer volume of \ce{H2O} with a thickness of \SI{10}{\cm}, contained within a \SI{0.64}{\cm} thick steel tank. 
Electrons that escape the central volume will still produce Cherenkov light  in this ``tail-catcher'' region, allowing a more complete integration of the total electron energy. 
Twelve Hamamatsu R5912-100 8 inch PMTs, immersed in the \ce{H2O}, will view the fiducial volume from above. 

To compensate for the absence of \num{4\pi} photocathode coverage, the inner walls of the steel tank will be covered in reflective Tyvek\textsuperscript{\textregistered}. 
Even commercially available Tyvek has favorable optical properties~\cite{Bugg:2013ica}. 
With this reflector, light collection within the \ce{D2O} volume will remain relatively uniform, but well-defined Cherenkov rings from, e.g., cosmic-ray backgrounds cannot be reconstructed.
Energy and timing cuts will therefore be the primary means of background discrimination. 

Outside the steel tank that supports the PMTs and encloses the tail catcher, \SI{5.08}{\cm} of lead shielding and two layers of \SI{2.54}{\cm} thick plastic scintillator panels read out via wavelength-shifting fibers~\cite{Bugg:2013ica} will mitigate external backgrounds due to beam-related neutrons, radioactivity in the hall, and cosmic rays. 
Additional detector designs are under consideration to increase the photon yield pending the availability of additional PMTs.

\section{Performance Goals}
\label{sec:performance}

The energy calibration will be established with Michel electrons from the decay at rest of cosmic-ray muons that stop in the fiducial volume.
These electrons have a well-characterized energy spectrum extending to about \SI{53}{\MeV}~\cite{Michel_1950,PhysRevLett.94.101805}, which overlaps with the electron energy spectrum from the \ce{\nu_{e} + d} interaction.
This technique is used by many collaborations; see for example Abe et al. ~\cite{PhysRevD.97.072001}. 
A system of light-emitting diode flashers, mounted inside the steel tank, will monitor the homogeneity and stability of the detector response.
This system will provide the capability to incorporate an understanding of photomultiplier gains and the transparency of the water and the acrylic volumes into the modeled detector response.
\begin{figure}
     \centering
         \includegraphics[width=0.6\textwidth]{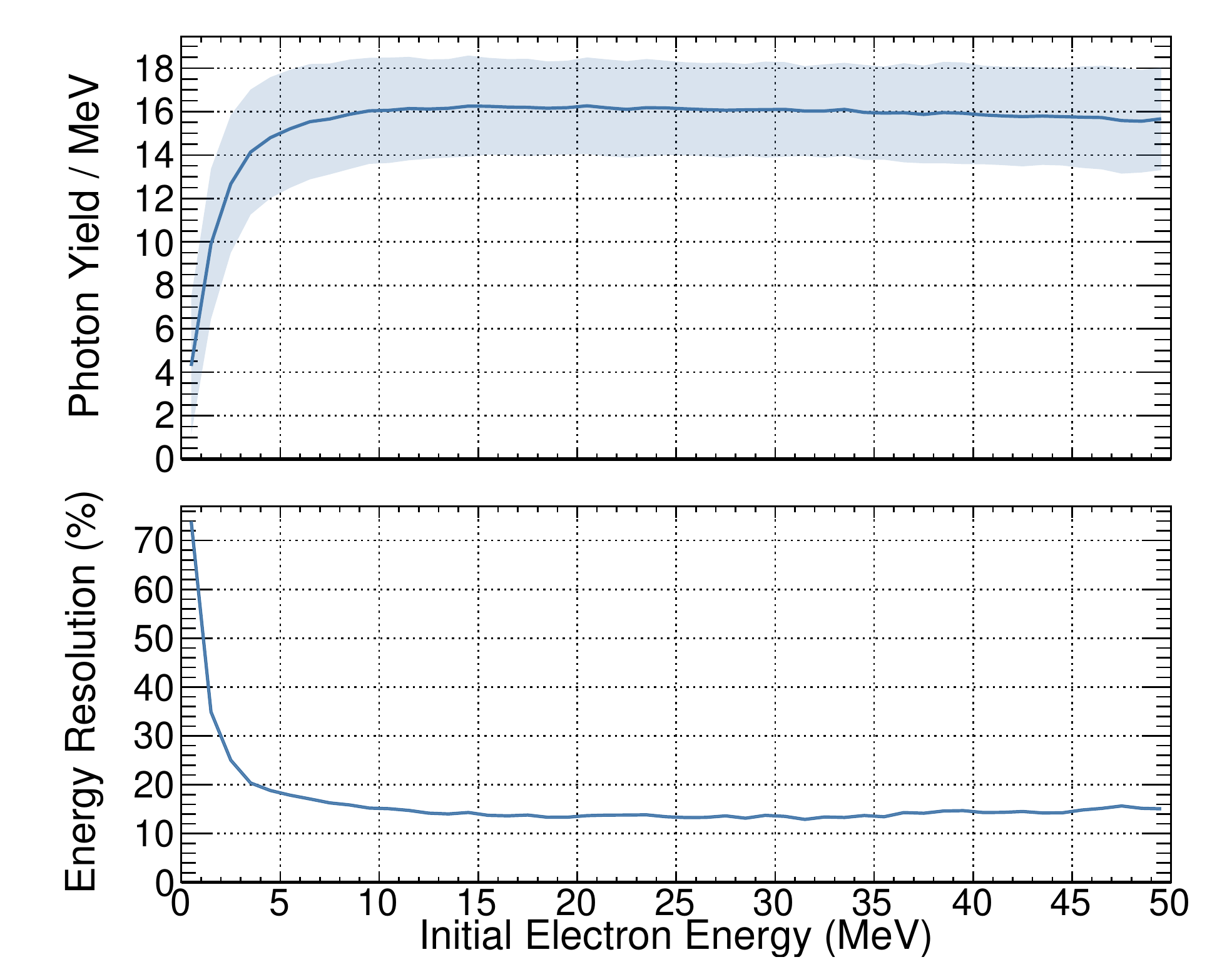}
        \caption{Expected detected photon yield (top) and fractional energy resolution ($\sigma/E_i$, bottom) for electrons, as a function of initial electron energy. The top panel shows a statistical $\pm 1\sigma$ band around the mean.}
        \label{fig:frac-eres}
\end{figure}

To evaluate the expected performance of the heavy-water detector, a \textsc{Geant4} simulation of the detector geometry discussed in Section~\ref{sec:design-fundamentals} was developed.
The simulation places the detector inside a mock-up of Neutrino Alley, including the \SI{2.4}{\m} of concrete overburden for cosmic-ray background predictions.

This detector simulation was used to predict both the expected energy resolution and the detected photon yield, as well as to evaluate the predicted spectra from various signals and backgrounds of interest, as discussed in Section~\ref{sec:sig_and_bkg}.
In the simulation, the reflectivity of the Tyvek coating was assumed to be \SI{97}{\percent}~\cite{6168236}.
The wavelength dependence of the PMT quantum efficiency~\cite{bib:hamamatsu} was also incorporated.

Figure~\ref{fig:frac-eres} shows the expected energy resolution and detected photon yield of the heavy-water detector.
For this \textsc{Geant4} simulation study, electrons (representing the detectable product of charged-current interactions) were generated isotropically within the fiducial volume, with known initial energies ranging uniformly from \SIrange[range-phrase=--, range-units=single]{1}{55}{\MeV}. 
The energy was then reconstructed based on the number of photoelectrons recorded by the simulated high-quantum-efficiency R5912-100 PMTs.
Within a narrow range of initial energies, the response peak was fit to a Gaussian with an exponential tail~\cite{1603.08591}, and the standard deviation was taken as the energy resolution. 
An expected photon yield of \SI{16}{photons\per\MeV} should provide sufficient energy resolution (roughly \SI{15}{\percent} above \SI{30}{\MeV}) to separate the charged-current \ce{\nu_{e} + d} signal from the expected backgrounds discussed below.
Note that even with the Tyvek reflector the light collection uniformity is not uniform throughout the detector, resulting in the leveling off of the energy resolution above roughly \SI{10}{\MeV}, largely driven by the roughly \SI{11}{\percent} non-linearity in the detected photon yield between the top and bottom of the detector.
Strategies to mitigate this effect are under investigation.
While the detected photon yield was found to depend on the assumed reflectivity of the Tyvek reflector, the assumed reflectivity was found to have a minimal effect on the reconstructed signal and background spectra.

\section{Signal and Background Predictions}
\label{sec:sig_and_bkg}

\subsection{Signal Prediction}
\begin{figure}
    \centerline{
        \includegraphics[width=0.545\textwidth,page=1]{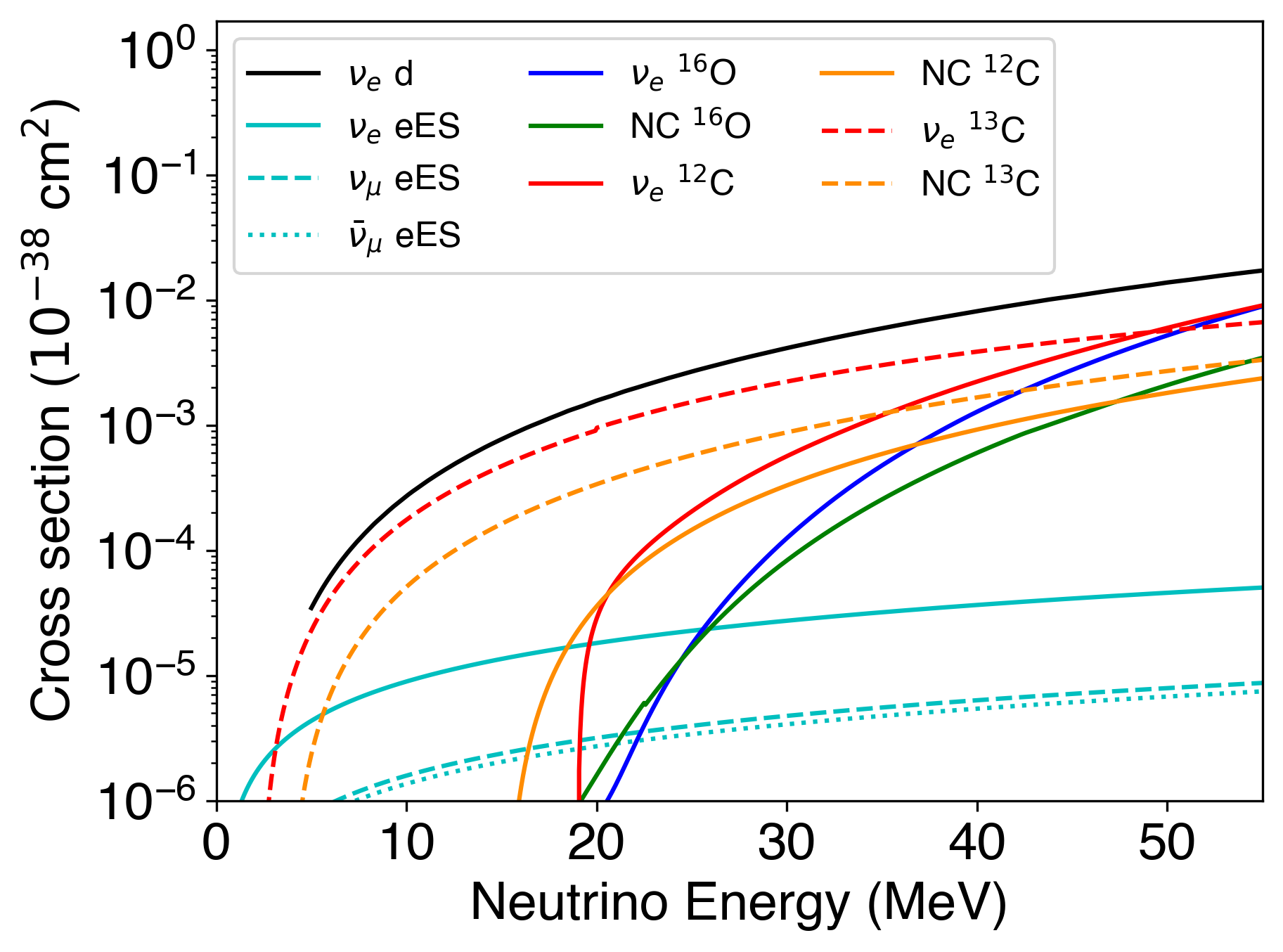}
    }
    \caption{Cross sections for neutrino interactions with heavy-water detector materials. `eES' refers to neutrino-electron elastic scattering.}
    \label{fig:xscns}
\end{figure}

The \textsc{Geant4} simulation described in Section~\ref{sec:performance} was used to predict the observed \ce{\nu_{e} + d} signal.
The charged-current \ce{\nu_{e} + d} cross section~\cite{Nakamura:2002jg, dxsec}, see Figure~\ref{fig:xscns}, was folded with the estimated neutrino flux~\cite{Rapp:2019vnv} at the location of the \ce{D2O} detector to form the true charged-current-induced electron spectrum (Figure~\ref{fig:rates}).
The response matrix obtained in Section~\ref{sec:performance} was then folded with this distribution to produce the expected observed spectrum after running for \SI{2}{SNS}-years (where an \si{SNS}-year is defined as \SI{5000}{hours} of operation at \SI{1.4}{\mega\watt}) as seen in Figure~\ref{fig:rates_reco}.

\subsection{Background Predictions}
\label{sec:bkg}

While beam-related neutrons are a significant background for \cevns searches at the SNS, they are not expected to be a major background for the heavy-water detector, as their interactions produce signals below the Cherenkov threshold in \ce{D2O}.
Data from other COHERENT detectors do not indicate the presence of a significant delayed neutron flux (arriving after the \SI{800}{\ns} SNS beam pulse) in Neutrino Alley~\cite{Akimov:2017ade, Akimov:2020pdx, Akimov:2019rhz}.
This finding further mitigates any potential background from beam-related neutrons, as the $\nu_{e}$ neutrino signal of interest will arrive delayed relative to the beam time profile.
Any potential background from delayed thermal neutron captures in the tail-catcher region will be removed with an energy threshold.
Nevertheless, the beam-related neutron flux will be monitored by the existing Multiplicity And Recoil Spectrometer (MARS) detector~\cite{ROECKER201621} in Neutrino Alley.
\begin{figure}
    \centerline{
        \includegraphics[width=0.495\textwidth,page=1]{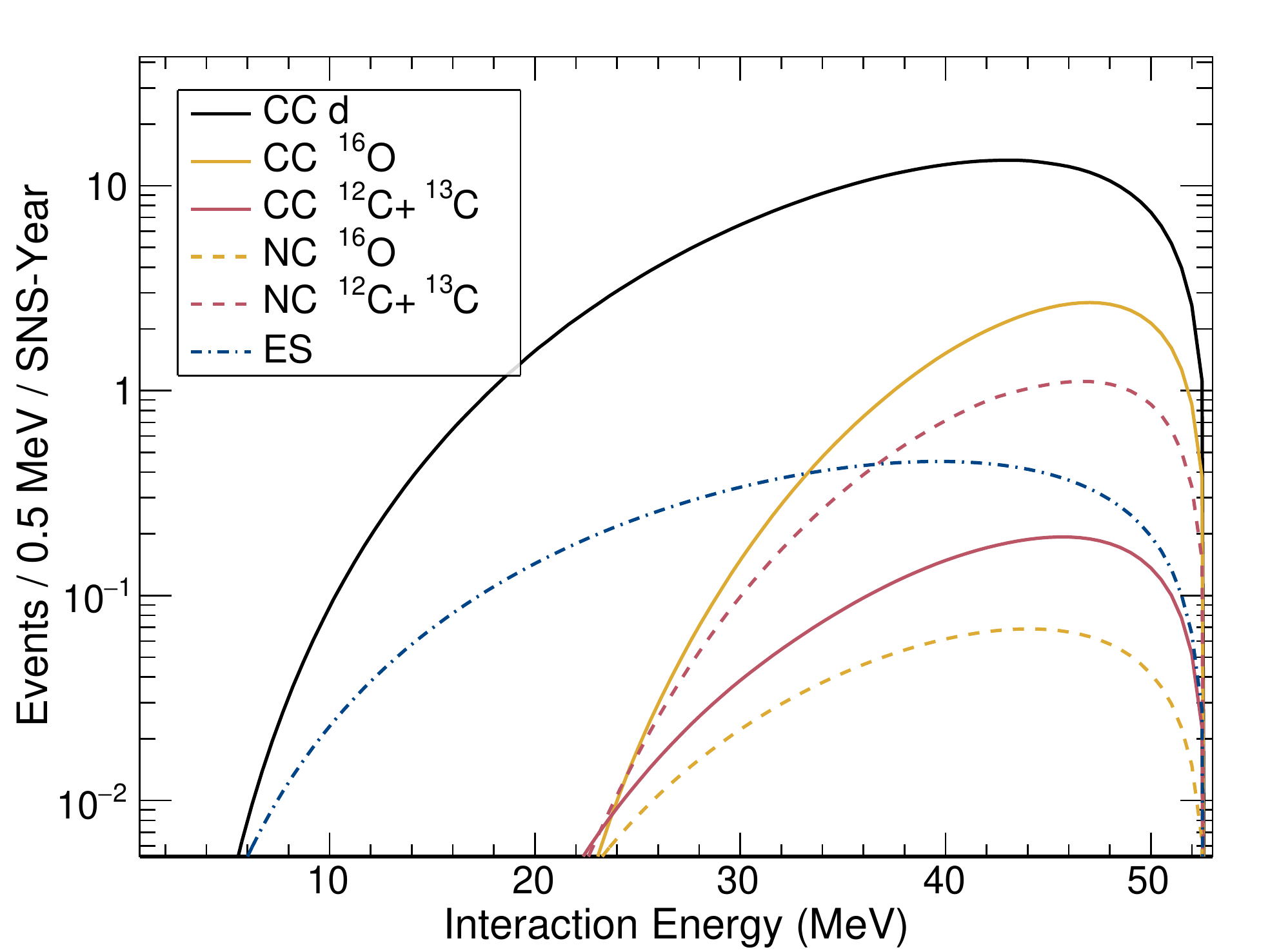}
        \includegraphics[width=0.495\textwidth,page=2]{Figures/snowglobes_plot_v3.pdf}
    }
    \caption{\textit{Left}: Estimated count rates for various neutrino interactions per SNS beam-year as predicted by SNOwGLoBES~\cite{snowglobes}, as a function of neutrino energy. Note that neutral-current interactions result in deexcitation gammas, for which only a fraction of the energy will be observed. \textit{Right}: Expected count rates for charged-current neutrino interactions in the \ce{D2O} detector for dominant interactions.
    Smearing from imperfect energy resolution was not included in this plot. Smeared spectra are presented in Figure~\ref{fig:rates_reco}.}
    \label{fig:rates}
\end{figure}

Neutrino-induced backgrounds are estimated using the SNOwGLobES~\cite{snowglobes} package, using cross sections shown in Fig.~\ref{fig:xscns}.
Charged- and neutral-current interactions with oxygen can occur in the fiducial volume or in the tail-catcher region.  
The $\nu_e$ on \ce{{}^{16}O} interaction~\cite{Haxton:1988mw} is also a signal of interest, see Section~\ref{sec:nu-d2o}, and may also be measured with a separate \ce{H2O} run of the detector.  
The neutral-current excitation of oxygen~\cite{Langanke:1995he} can produce deexcitation gammas, but with energies less than roughly \SI{8}{\MeV}, and for which Compton scatters produce yet smaller event energies---so these should be a negligible background.

Although charged- and neutral-current interactions with the carbon atoms~\cite{Auerbach:2001hz, Armbruster:1998gk} in the acrylic are possible, the rates are negligible because of the low acrylic mass in the detector.
Any charged-current events from carbon that do occur may also be removed with an energy-threshold cut.
\begin{figure}
    \centerline{
        \includegraphics[width=0.495\textwidth]{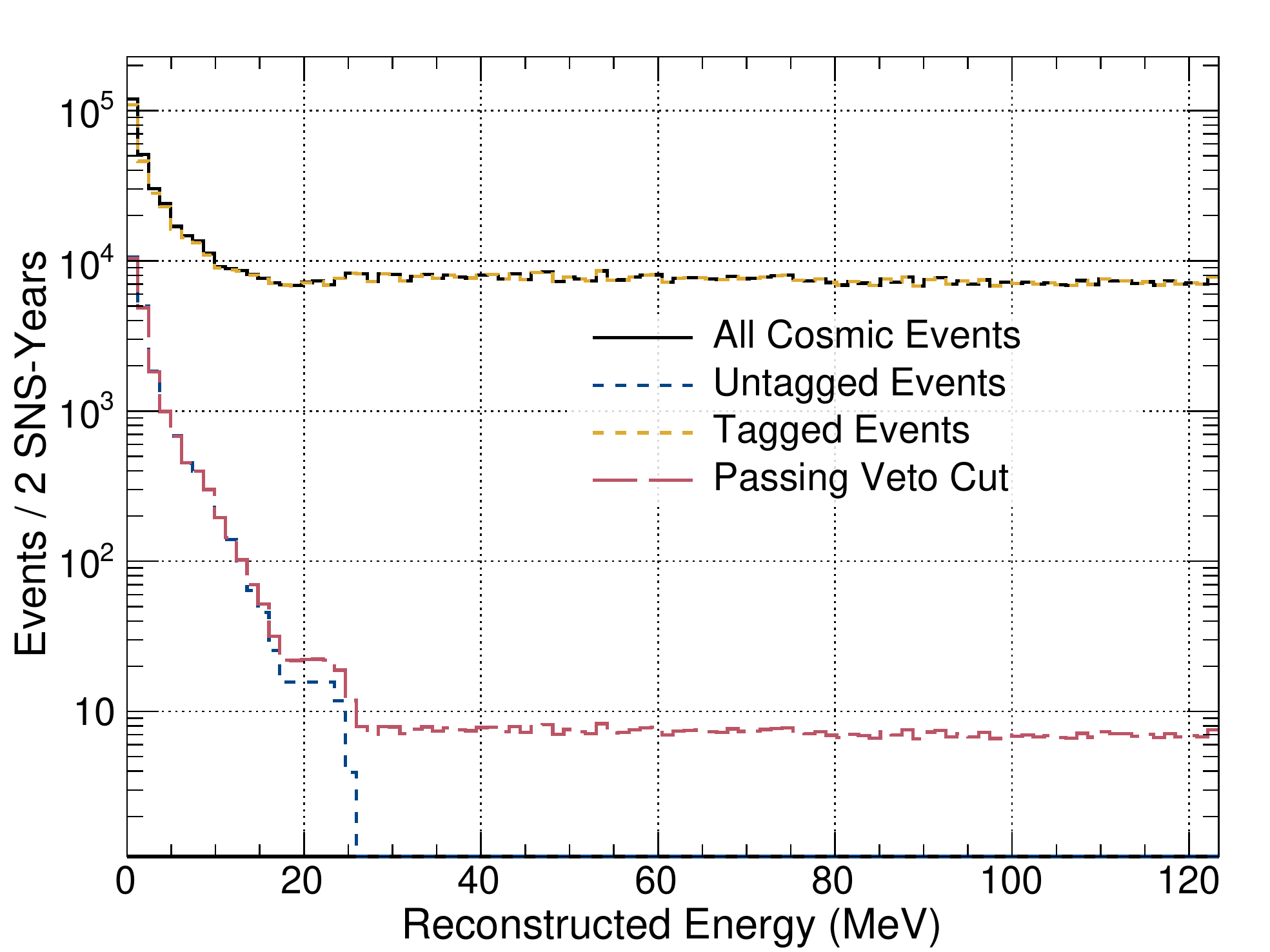}
        \includegraphics[width=0.495\textwidth]{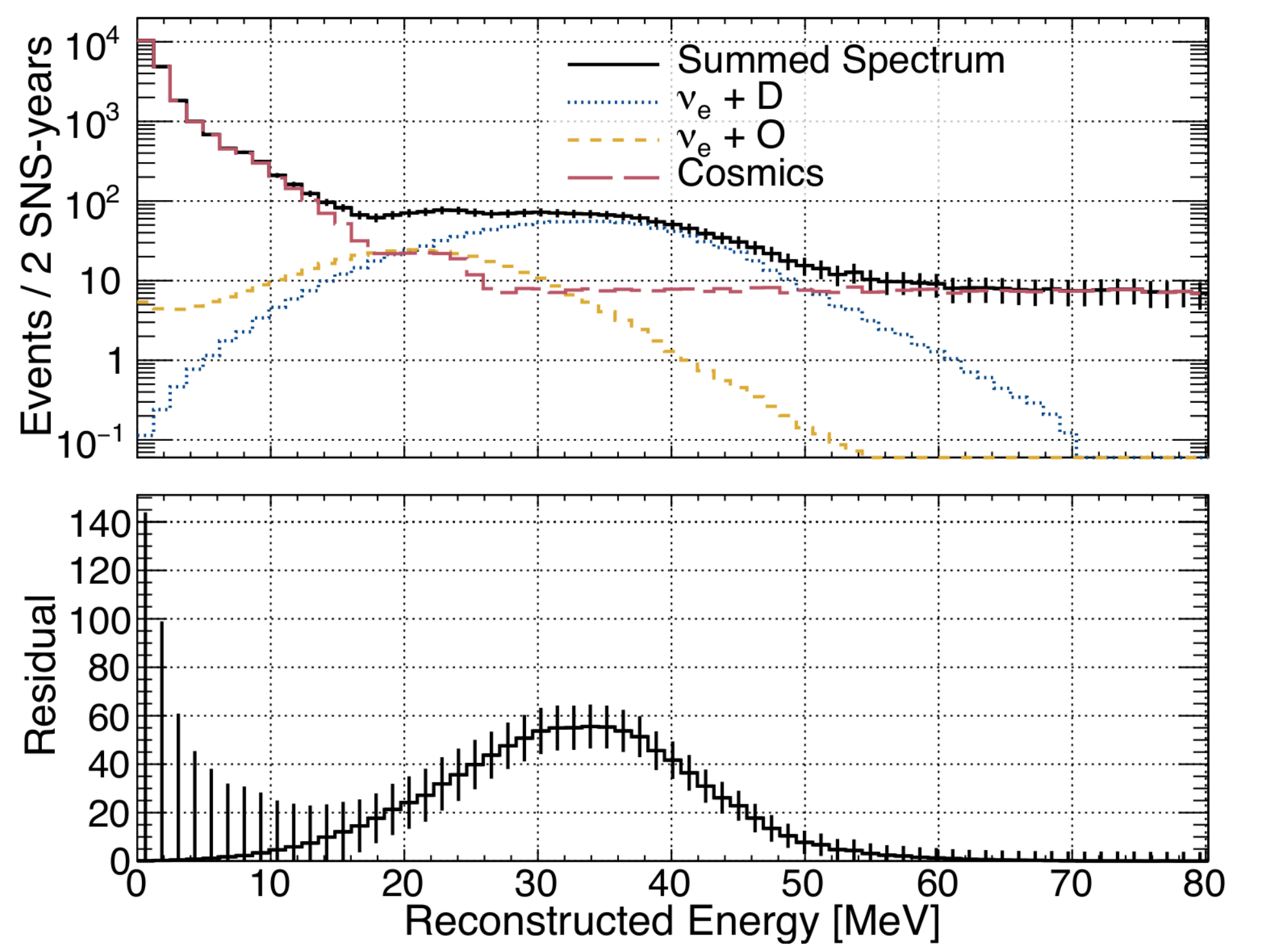}
    }
    \caption{Expected event rates in the heavy-water detector.
    \textit{Left:} Expected background from cosmic rays as predicted with the CRY~\cite{4437209cry} module and simulated with the \textsc{Geant4} simulation described in Section~\ref{sec:performance}, with and without a muon veto as discussed in Section~\ref{sec:bkg}.
    \textit{Top right}: Simulated signal and background energy spectra as reconstructed with the heavy-water detector. 
    \textit{Bottom right}: The background-subtracted \ce{\nu_{e} + d} spectrum with statistical errors. 
    As discussed in the text, an energy threshold above roughly \SI{20}{\MeV} will remove most of the background contamination from the expected deuteron signal. 
    Smearing from imperfect energy resolution is included here.
    }
    \label{fig:rates_reco}
\end{figure}
Elastic scatters of neutrinos  on electrons occur with well-known cross sections, but are a tiny contribution to the event rate.
Interaction rates as a function of neutrino energy are shown in Fig.~\ref{fig:rates}.
Note that neutrino-induced neutrons produced in the detector shielding are not anticipated as a major background as they are typically a few \si{MeV} in energy and well below the energy region of interest.

As the dominant neutrino-produced beam-related background, the true charged-current-on-\ce{O} energy spectrum was folded with the response matrix formed from the simulation discussed in Section~\ref{sec:performance} to produce the spectrum shown in Figure~\ref{fig:rates_reco}.
Because the energy spectrum of electrons produced from charged-current interactions with oxygen peaks approximately \SI{18}{\MeV} lower in energy, they can be suppressed with an energy threshold as seen in Figs.~\ref{fig:rates} and \ref{fig:rates_reco}.

The most significant expected background is cosmic-ray products that lie above the Cherenkov threshold in \ce{D2O}, especially given the expected lack of angular reconstruction capabilities of the detector.
The CRY~\cite{4437209cry} cosmic-ray-generator package was used as input to the \textsc{Geant4} simulation described in Section~\ref{sec:performance} to predict the cosmic-ray-induced backgrounds for the heavy-water detector.
Cosmic rays were generated at the surface and transported through the \SI{2.4}{\m} of concrete overburden above Neutrino Alley.
The thickness of the concrete overburden was tuned to match the observed muon rate measured with a muon telescope in Neutrino Alley.
It is important to note that while the cosmogenic backgrounds have been predicted with simulations here, in reality they will be well measured in between SNS beam pulses and our evaluation here is not sensitive to the details of the simulation.
Nevertheless, the predicted cosmic rate will be compared to the measured rate by a light-water detector operating at a similar overburden to Neutrino Alley which is being used for PMT testing prior to construction of the first \ce{D2O} detector module.

There are two classes of cosmic-ray-induced background events: tagged and untagged.
Tagged events are those that deposit more than \SI{5}{\MeV} in the muon veto.
With the high light-collection efficiencies possible in veto panels~\cite{Bugg:2013ica} and good hermiticity, muons will be tagged with an efficiency close to \SI{99.9}{\percent}.
Even with the beam duty factor (\num{6e-4}), this high efficiency is needed because of the shallow overburden in Neutrino Alley.
As seen in Figure~\ref{fig:rates_reco}, these backgrounds, and in particular higher-energy cosmic rays with a reconstructed energy of $\gtrsim\SI{20}{\MeV}$, can be largely removed with the inclusion of a two-layer muon veto.
Deadtime concerns due to the cosmic-ray flux are anticipated to be on the order of a few percent based on the measured muon flux in Neutrino Alley.

\section{Estimated Rates and Sensitivity}
\label{sec:rates}

The results from the \textsc{Geant4} simulations described in Section~\ref{sec:performance} and \ref{sec:sig_and_bkg} were used to predict the rates from the charged-current \ce{\nu_{e} + d} signal and the dominant expected background spectra and rates.
Events are considered `detected' if they pass a \SI{2}{PE} cut.
As seen in Figure~\ref{fig:rates} and Figure~\ref{fig:rates_reco}, most expected backgrounds reside at a lower energy than the charged-current deuterium signal.
This result provides the opportunity to minimize the background contamination of the signal with an energy acceptance window.
While a full likelihood fit would maximize the expected sensitivity to the deuterium signal, a simple one-binned counting experiment of events within an energy window is sufficient to conservatively estimate the statistical precision possible with the heavy-water detector.

The statistical precision is given by
\begin{equation} \label{eq:prec}
    \alpha \equiv \frac{\sigma}{N} = \frac{\sqrt{N + 2B}}{N},
\end{equation}
where $\sigma$ is the statistical uncertainty of the number of signal events $N$, and $B$ is the expected number of background events.
In this case, an additional scale factor of 2 in the numerator follows the pessimistic assumption that the backgrounds will not be well measured.
In reality, the cosmic background will be well measured between proton pulses, reducing the factor of 2 and increasing the signal significance.
The background from oxygen charged-current events will occur in tandem with the beam and will need either to be measured with a separate \ce{H_{2}O} fill of the detector or to be modeled by simulation and included in a simultaneous fit, but the uncertainty from this background is anticipated to be subdominant.

\begin{figure}
    \centerline{\includegraphics[width=0.6\textwidth]{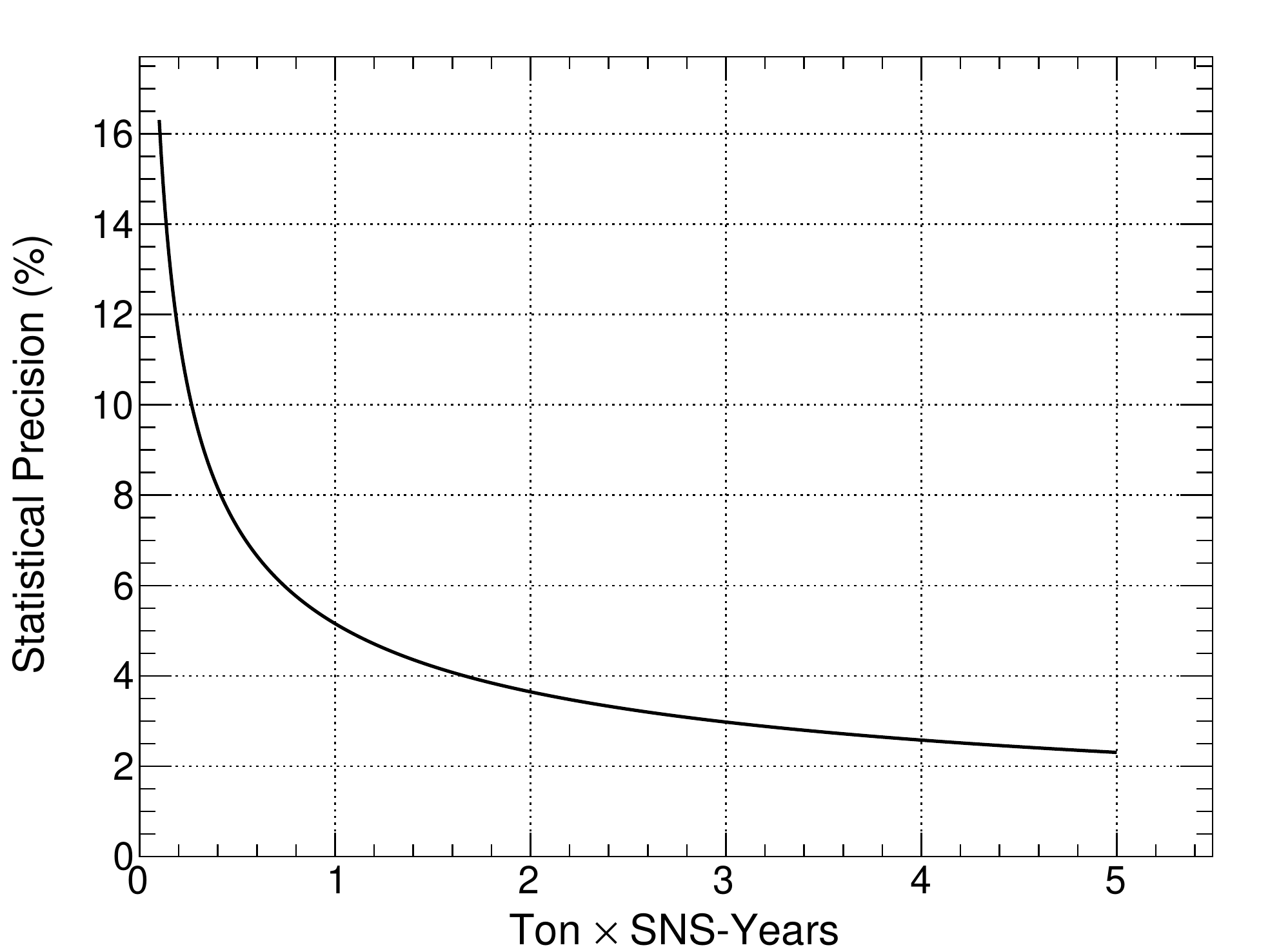}}
    \caption{Expected statistical uncertainty following Eq.~\ref{eq:prec} of the number of charged-current-deuterium events as a function of mass and time. The SNS operates for \SI{5000}{h} every year at \SI{1.4}{\mega\watt}.}
    \label{fig:prec_over_time}
\end{figure}

The expected spectra due to the charged-current-deuterium signal and the dominant backgrounds may be seen in Figure~\ref{fig:rates_reco}.
An optimal energy window was found to minimize the statistical uncertainty of the background-subtracted number of signal events.
For the heavy-water detector, an energy acceptance window of greater than \SI{22}{\MeV} and less than \SI{68}{\MeV} minimized the statistical uncertainty of a one-bin counting experiment, with an expected precision of \SI{4.7}{\percent} in \SI{2}{SNS}-years of running time, as seen in Figure~\ref{fig:prec_over_time}.

Table~\ref{tab:evt_rates} summarizes the expected event rates in the single-module heavy-water detector following \SI{2}{SNS}-years of running both with and without the energy cut discussed earlier.
As seen in the table, a total of \SI{1070}{signal\,events} are expected in \SI{2}{years} of run time, with \num{910} of those events occurring within the optimal energy range for a counting experiment.

\begin{table}[b]
    \centerline{
    \begin{tabular}{r r r}
                      & \textbf{Total Events} & \textbf{Events in Region of Interest} \\
    \toprule
    $\bm{\nu_{e}+ d}$ & 1070                  & 910 \\
    \midrule
    $\bm{\nu_{e}+ O}$ & 390                   & 160 \\
    \textbf{Cosmics}  & 21150                 & 315 \\
    \bottomrule
    \end{tabular}
    }
    \caption{Expected signal and major-background event rates in the single-module heavy-water detector following \SI{2}{SNS}-years of running at \SI{1.4}{\MW} and a duty factor of \num{6e-4}. Energy, timing, and veto cuts will remove much of the backgrounds for the \ce{\nu_{e} + d} signal as discussed in the text.
    An optimal energy region of interest from \SIrange[range-phrase=--, range-units=single]{22}{68}{\MeV} will produce an expected statistical precision for a simple event-counting analysis of \SI{4.7}{\percent}.}
    \label{tab:evt_rates}
\end{table}
\section{Potential Physics Impact}
\label{sec:phys_impact}

The Standard Model predicts that the \cevns cross section is approximately proportional to the square of the number of neutrons in the nucleus ($\propto N^{2}$).
One of the main goals of the COHERENT experiment is to verify this $N^{2}$ prediction by using a suite of detectors with different nuclei.
In such a comparison, the SNS neutrino flux represents a common factor among the detectors and drops out in the ratio.  Furthermore, physics studies that depend primarily on recoil spectrum shape will not be especially sensitive to absolute flux uncertainty.
However, the neutrino flux is a significant source of uncertainty for any measurement involving an absolute cross section.
In particular, the current and future measurements of inelastic cross sections at the SNS~\cite{Akimov:2018ghi} will benefit from an improved understanding the neutrino flux in neutrino alley.
An improved measurement of these cross sections will have wide-ranging implications ranging from our understanding of the DUNE supernova signature~\cite{2020arXiv200806647D} to our understanding of $r$-process nucleosynthesis in supernova explosions through neutrino-induced neutron production~\cite{Qian:1996db}, and the interpretation of the HALO supernova signature~\cite{Duba_2008}, to improved constraints on $g_{A}$ quenching~\cite{Engel_2017, Menendez:2011qq}.
A reduced neutrino flux uncertainty will also allow neutrino experiments at the SNS to inform our understanding of nuclear structure~\cite{Patton-NeutronDensity-2012, Amanik_2008}.

Let us illustrate the effects of a reduced neutrino flux uncertainty with one example.
As \cevns is a neutral-current process, the cross section is flavor-independent in the Standard Model, with small differences from radiative corrections~\cite{Sakakibara:1979rc, Sehgal:1985iu, Tomalak:2020zfh}.
So-called non-standard neutrino interactions (NSI), which are crucial to understanding the mass-ordering determination from long-baseline neutrino-oscillation experiments~\cite{ColomaSchwetz-MassOrderingDegeneracy:2016, coloma:2017, Coloma-LMADark:2017}, can manifest as a flavor dependence of the \cevns cross section.
The precise timing of the SNS beam allows for flavor separation of the incident neutrinos, providing an opportunity to search for a flavor dependence of the \cevns cross section.
\begin{figure}
    \centerline{
        \includegraphics[width=0.495\textwidth]{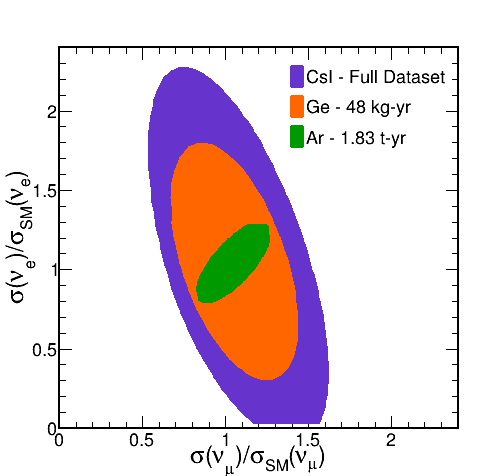}
        \includegraphics[width=0.495\textwidth]{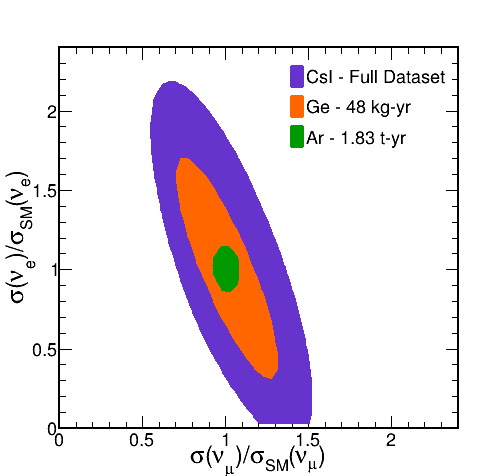}
    }
    \caption{Predicted sensitivity to deviations from the Standard Model \cevns cross section for $\nu_{\mu}$ and $\nu_{e}$ neutrinos. The ``\ce{CsI}--Full Dataset'' regions correspond to the full \ce{CsI} dataset from the COHERENT experiment with roughly twice the accumulated statistics of the initial result~\cite{magcevns_dan_2020}. Any deviation from $(1,1)$ could be evidence of beyond-Standard Model physics. \textit{Left}: Current \SI{10}{\percent} neutrino-flux uncertainty. \textit{Right}: Allowed regions assuming the SNS neutrino flux is measured with a precision matching the current \SI{3}{\percent} theoretical uncertainty of the \ce{\nu_{e} + d} cross section.}
    \label{fig:nsi}
\end{figure}

Taking advantage of this flavor separation, Figure~\ref{fig:nsi} shows the expected constraints on the $\nu_{\mu}$ and the $\nu_{e}$ \cevns cross sections relative to the Standard Model predictions.
Shown in the left panel are the expected constraints on the \cevns cross section for electron- and muon-flavored neutrinos relative to their Standard-Model predictions with the current \SI{10}{\percent} uncertainty of the incident neutrino flux.
The right panel shows the improved constraints assuming the flux uncertainty can be reduced to the theoretical \SI{3}{\percent} uncertainty of the \ce{\nu_{e} + d} cross section following deployment of the second heavy-water detector module at the SNS, as well as the next generation of COHERENT CEvNS detectors discussed in Section~\ref{sec:intro}.
The improved knowledge of the neutrino flux enabled by a heavy-water detector at the SNS would reduce the allowed parameter space by roughly a factor of two.
Any deviations from the Standard Model prediction, and in particular any flavor dependence of the cross section, would be an indication of physics beyond the Standard Model.

\section{Conclusions}
One of the dominant systematic uncertainties for neutrino detectors operating at the pion decay-at-rest neutrino source at the ORNL SNS is the \SI{10}{\percent} uncertainty of the incident neutrino flux.
A heavy-water Cherenkov detector has been designed to capitalize on the small (\SIrange[range-phrase=--, range-units=single]{2}{3}{\percent}) theoretical uncertainty of the charged-current \ce{\nu_{e} + d} interaction.
The detector will fit in the available space within Neutrino Alley and make a direct measurement of the SNS neutrino flux with a statistical uncertainty of \SI{4.7}{\percent} in \SI{2}{SNS}-years.
This detector is the first of a two-module design that will enable the COHERENT experiment to normalize the neutrino flux from the SNS, ultimately to \SI{3}{\percent} precision, allowing a corresponding increase in sensitivity to new physics beyond the Standard Model.

As a case study in the increased physics sensitivity capable of being probed with a reduced neutrino flux uncertainty at the SNS, the allowed region for a flavor dependence of the \cevns cross section was studied.
The allowed region was shown to shrink by roughly a factor of 2, assuming a reduction in the flux uncertainty from the current \SI{10}{\percent} to an uncertainty of \SI{3}{\percent}, matching the theoretical uncertainty of the \ce{\nu_{e} + d} cross section.
Following this initial program to calibrate the neutrino flux for \SI{1}{\GeV} protons at the SNS, the detector will also be used to calibrate the flux at \SI{1.3}{\GeV} following the SNS Proton Power Upgrade~\cite{bib_ppu}. 
It may later be moved to the ORNL SNS Second Target Station~\cite{osti_1185891} to further support future neutrino efforts at the SNS.

\acknowledgments
The COHERENT collaboration acknowledges the resources generously provided by the Spallation Neutron Source, a DOE Office of Science User Facility operated by the Oak Ridge National Laboratory. 
This work was supported by the US Department of Energy (DOE), Office of Science, Office of High Energy Physics and Office of Nuclear Physics; the National Science Foundation; the Consortium for Nonproliferation Enabling Capabilities; the Institute for Basic Science (Korea, grant no.\,IBS-R017-G1-2019-a00); the Ministry of Science and Higher Education of the Russian Federation (Project ``Fundamental properties of elementary particles and cosmology'' No. 0723-2020-0041); the Russian Foundation for Basic Research (Project 20-02-00670\_a); and the US DOE Office of Science Graduate Student Research (SCGSR) program, administered for DOE by the Oak Ridge Institute for Science and Education which is in turn managed by Oak Ridge Associated Universities. 
Sandia National Laboratories is a multi-mission laboratory managed and operated by National Technology and Engineering Solutions of Sandia LLC, a wholly owned subsidiary of Honeywell International Inc., for the U.S. Department of Energy's National Nuclear Security Administration under contract DE-NA0003525. 
The Triangle Universities Nuclear Laboratory is supported by the U.S. Department of Energy under grant DE-FG02-97ER41033. 
Laboratory Directed Research and Development funds from Oak Ridge National Laboratory also supported this project. 
This research used the Oak Ridge Leadership Computing Facility, which is a DOE Office of Science User Facility.

\bibliographystyle{vitae}
\bibliography{D2O_Concept_v2}

\end{document}